\documentclass[ amsmath,amssymb,aps,prl,preprint]{revtex4-1}
\usepackage{appendix}
\usepackage{graphicx}
\usepackage{dcolumn}
\usepackage{bm}
\usepackage{hyperref}
\usepackage[mathlines]{lineno}
\DeclareGraphicsExtensions{.pdf,.PDF,png,.PNG}
\usepackage{xspace} 
\usepackage{soul}
\makeatletter
\gdef\@ptsize{0} 
\makeatother
\makeatletter
\usepackage{suffix}
\let\latex@section\section
\WithSuffix\def\section*{\secdef\my@section{\latex@section*}}
\def\my@section[#1]#2{}
\makeatother
\usepackage{xcolor}
\usepackage{algorithm}
\usepackage{algpseudocode}
\usepackage{amsmath}
\usepackage{longtable} 

\usepackage{xr}
\begin{document}
 \title{Facilitating a 3D granular flow with an obstruction}
 \author{Abhijit Sinha, Shankar Ghosh}
\affiliation{Department of Condensed Matter Physics and Materials Science, Tata Institute of Fundamental Research, Homi Bhabha Road, Mumbai 400-005, India}%
\author{Jackson Diodati, Narayanan Menon, Shubha Tewari}
\affiliation{Department of Physics, University of Massachusetts, Amherst, MA 01003, USA.}
\date{\today}




\begin{abstract}
Ensuring a smooth rate of efflux of particles from an outlet without unpredictable clogging events is crucial in processing powders and grains. We show by experiments and simulations that an obstacle placed near the outlet can greatly suppress clog formation in a 3-dimensional granular flow; this counterintuitive phenomenon had previously been demonstrated in 2-dimensional systems. Remarkably, clog suppression is very effective even when the obstacle cross-section is comparable to that of a single grain, and is only a small fraction of the outlet area.  
Data from two dissimilar obstacle shapes indicate that the underlying mechanism for clog suppression is geometric: while the magnitude of the effect can be affected by dynamical factors, the optimal location of the obstacle is  determined entirely by a simple geometric rule. An optimally-placed obstacle arranges for the most-probable clogging arch to be at a perilous and unstable spot, where the local flow geometry is at a point of expansion and a slight perturbation leaves the arch without downstream support. This geometric principle does not rely on previously conjectured dynamical mechanisms and admits generalization to other agent-based and particulate systems.
    
\end{abstract}
\maketitle

Dispensing granular materials such as pharmaceuticals, food powders, food grains, or construction material at a predictable rate from a hopper is a crucial step in industrial processes. This is typically achieved by gravity-driven efflux out of a hopper or a funnel, with the flow rate being controlled by varying the size of the outlet, $a$, relative to that of an individual grain, $d$, in a manner that is well-described by the empirical Beverloo relation, where the flow rate scales as  $  (a-kd)^{5/2}$. Here, $k$ is a constant that depends on particle shape and friction \cite{nedderman1992statics}.  However, when $a$ is reduced to achieve a slow flow rate, the flow is interrupted by intermittent clogging events associated with the formation of load-bearing jammed structures that block the outlet. As these clogs occur unpredictably, they have to be cleared either by operator intervention or by a periodic input of energy via mechanical vibration \cite{janda2009unjamming} or gas jets  \cite{zhu2020hopper}.

Placing obstacles near an exit to reduce congestion by slowing down exiting agents is known to be a successful method in fields like traffic management and panic evacuation\cite{helbing2000simulating,escobar2003architectural}.
Motivated by this analogy with pedestrian discharge from an enclosed space,  Zuriguel and coworkers \cite{zuriguel2011silo} demonstrated an ingenious, simple, and counterintuitive strategy to alleviate jamming by inserting an \textit{obstacle} near the outlet. They showed that the presence of an optimally located object changes the flow rate predicted above only slightly, while greatly suppressing the probability of clogging, without any intervention or energy input.  The suppression of clogging by an obstacle has since been heavily investigated by them and others in 2D experiments and simulations \cite{garcimartin2010shape,alonso2012bottlenecks,zuriguel2011silo,endo2017obstacle,harada2022silo,bick2021strategic}.  These studies have explored ways to both optimize the effect as well as to understand the underlying mechanism by which an obstacle actually facilitates a steady flow. Many of the proposed mechanisms elaborate on some aspect of the disruption of the flow by the obstacle. Some conjectures are that the flow slows down just upstream of the obstacle and then speeds up as it divides into two channels (`the waiting room effect') \cite{escobar2003architectural,alonso2012bottlenecks}; 
that particles scattering off the obstacle lead to horizontal velocity fluctuations that can break incipient arch formation \cite{gella2022dual}; that the flow downstream of the object is not laminar and leads to a lower packing fraction above the outlet \cite{gao2019understanding}.  Other possibilities are more probabilistic: there is a squared suppression of the clogging probability for arches to span simultaneously on both sides of the obstacle, and if the flow clogs on one side of the obstacle, mechanical noise from the flowing side causes it to break. 

Even if there is not a complete consensus on the underlying mechanism, the robust success of this stratagem of placing obstacles
in granular flows 
stimulated analogous studies in other problems of field-driven and activity-driven escape through a finite outlet, in settings ranging from sheep-herding and pedestrian traffic to microfluidics \cite{zuriguel2014clogging,patterson2017clogging}.  All of these settings,  however, are essentially 2-dimensional, and surprisingly, there are only very limited studies of the all-important case of flow from a 3-dimensional container.  In this article, we show definitively by both experiments and numerical simulations that the suppression of clogs by an obstacle also holds in 3D. Furthermore, we find that relatively small obstacles, comparable to the size of the particle, can be very effective in facilitating flow, producing up to an eight-fold suppression of clogging probabilities relative to a system without an obstacle, even though the area occluded by the obstacle is a \textit{small} fraction of the cross-section of the flow. Finally, in the 3D geometry we explore, clog suppression cannot be readily accounted for by the arguments advanced for explaining the effect in 2D. Instead, both experiments and simulations point to a geometric explanation for this effect, where the role of the obstacle is to place the most-likely clog at a location where it is mechanically weakest.


\begin{figure}[t]
    \includegraphics[width=.95\textwidth]{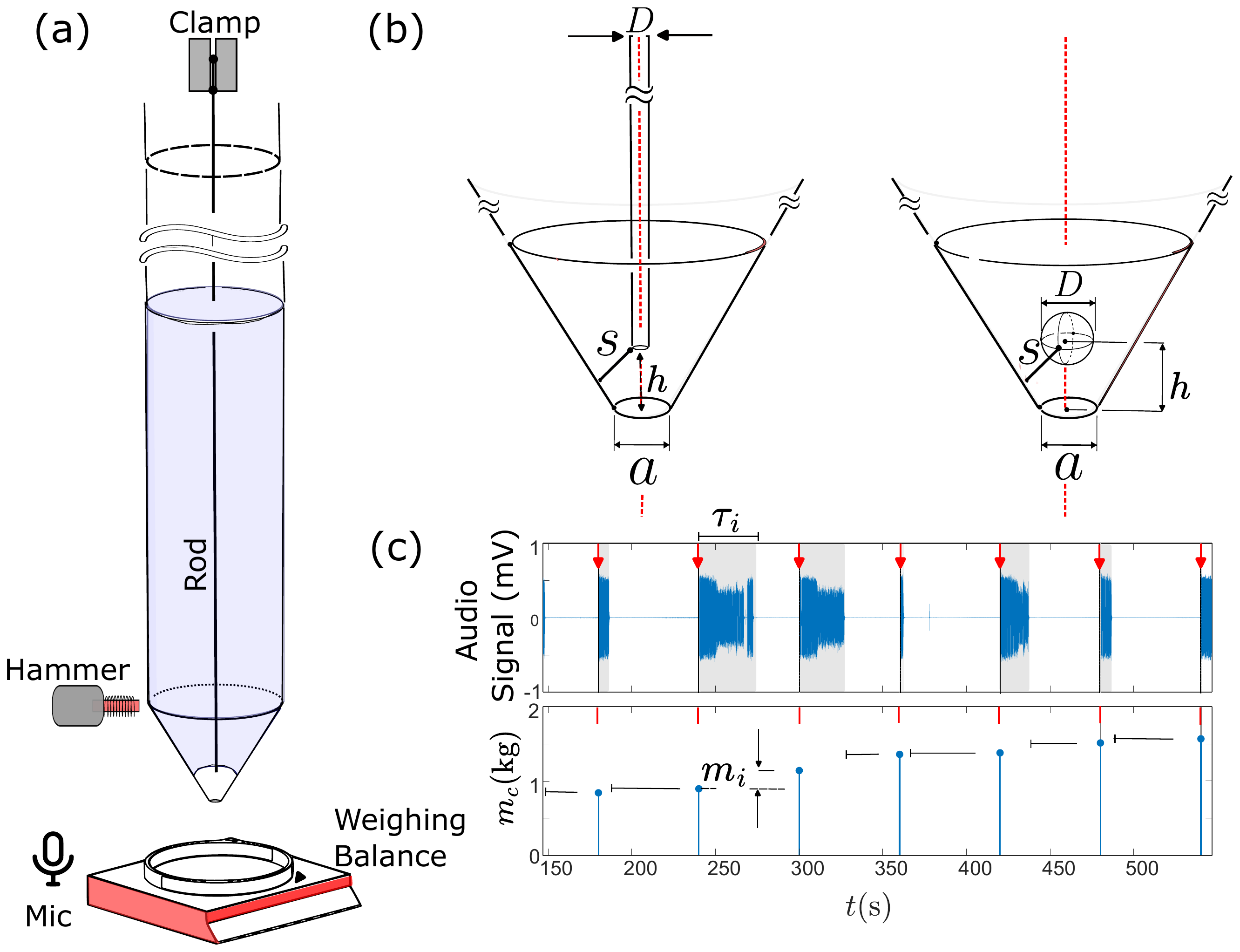}
    \caption{\textbf{ Flow geometry and measurements} (a) Setup of  experiment.
  (b) The left and right panels depict the region near the hopper exit in the experiment and simulation, respectively. The outlet diameter $a$ and obstacle diameter $D$ are indicated. In the experiment, $h$ represents the vertical height of the obstacle tip from the plane of the outlet, while in the simulation, $h$ is measured from the outlet to the  center of the sphere. In both the experiment and simulation, $s$ denotes the minimum distance from the obstacle surface to the nearest point on the hopper.
   (c) The time trace of the audio signal during discharge and the cumulative mass $m_c$ recorded by the balance are shown, with the duration of each clog $\tau_i$, and the incremental mass outflow $m_i$ indicated. }
        \label{fig:Fig1}
\end{figure}

\textbf{Flow geometry and grains} In both experiments and simulations, as shown in Fig. \ref{fig:Fig1} (a) spherical grains flow under gravity through a long cylindrical section ending in a conical funnel with interior angle $60 ^{\circ}$ and a circular outlet of diameter $a$ at its base. In the experiment, we use glass spheres with a diameter $d=2.5\pm 0.25$ mm 
whereas the simulation uses monodisperse spheres. The outlet size is chosen to produce relatively frequent clogs: in the experiments the outlet is held fixed at $a/d=3.2$, but we report simulation results from four different outlet sizes in this range of $a/d$.

\textbf{Obstacle geometry} In our simulations, the obstacle is a sphere of diameter $D$ on the axis of the hopper at height $h$ above the opening, as shown in Figure \ref{fig:Fig1} (b). This is therefore a faithful 3D representation of the geometry explored in previous 2D studies, exemplified by Ref. \cite{zuriguel2011silo}. In experiments, however, an obstacle requires a suspension mechanism such as a rigid rod. Our preliminary measurements, however, showed that the suspending rod itself played a remarkably effective role in suppressing clogging, thus all the experiments reported here are done with only a rigid glass or graphite rod of diameter $D$ placed with its tip at a height $h$ above the outlet (Fig. \ref{fig:Fig1} (b)). For most of the measurements reported here, we suspend the upper end from a string above the hopper. Potentially, this allows the rod to adjust its position, however, we observe almost no motion of the rod tip, and the rod appears to dynamically center itself in the flow.  In what follows, we report the effects of varying both the obstacle location $h$ and obstacle size $D$ (diameter of sphere or rod). 

    \textbf{Initial conditions and clogs} The simulations are run using the granular package within the LAMMPS\cite{lammps2022} environment, with Hertzian particle-particle and particle-wall interactions \cite{DEM}, a static coulomb friction coefficent $\mu$, and viscoelastic damping for normal and tangential relative displacements of grains \cite{silbert2001}.  Flows are initialized by 'pouring' $\sim 10^4$ grains into the closed hopper.  Once they settle into a static configuration, we open the hopper outlet, and grains flow out till they clog, as determined by a vanishing outflux of grains. To release the clog and re-initialize flow, we remove all grains within a fixed height above the outlet of the hopper and reintroduce these grains at the top of the hopper, and the simulation continues with a constant number of grains. In the experiments, too, the hopper is filled and allowed to drain under gravity, while always maintaining the height of the column well above the conical section.  To release clogs, we make a preliminary determination of the typical time to clog, $\langle\tau\rangle$, and then program a hammer powered by a push-pull solenoid to periodically strike the hopper. The time interval between successive hammer strikes is chosen to be much greater than $\langle\tau\rangle$, ensuring that the flow completely ceases before the next strike occurs.


\begin{figure}
    \includegraphics[width=.95\textwidth]{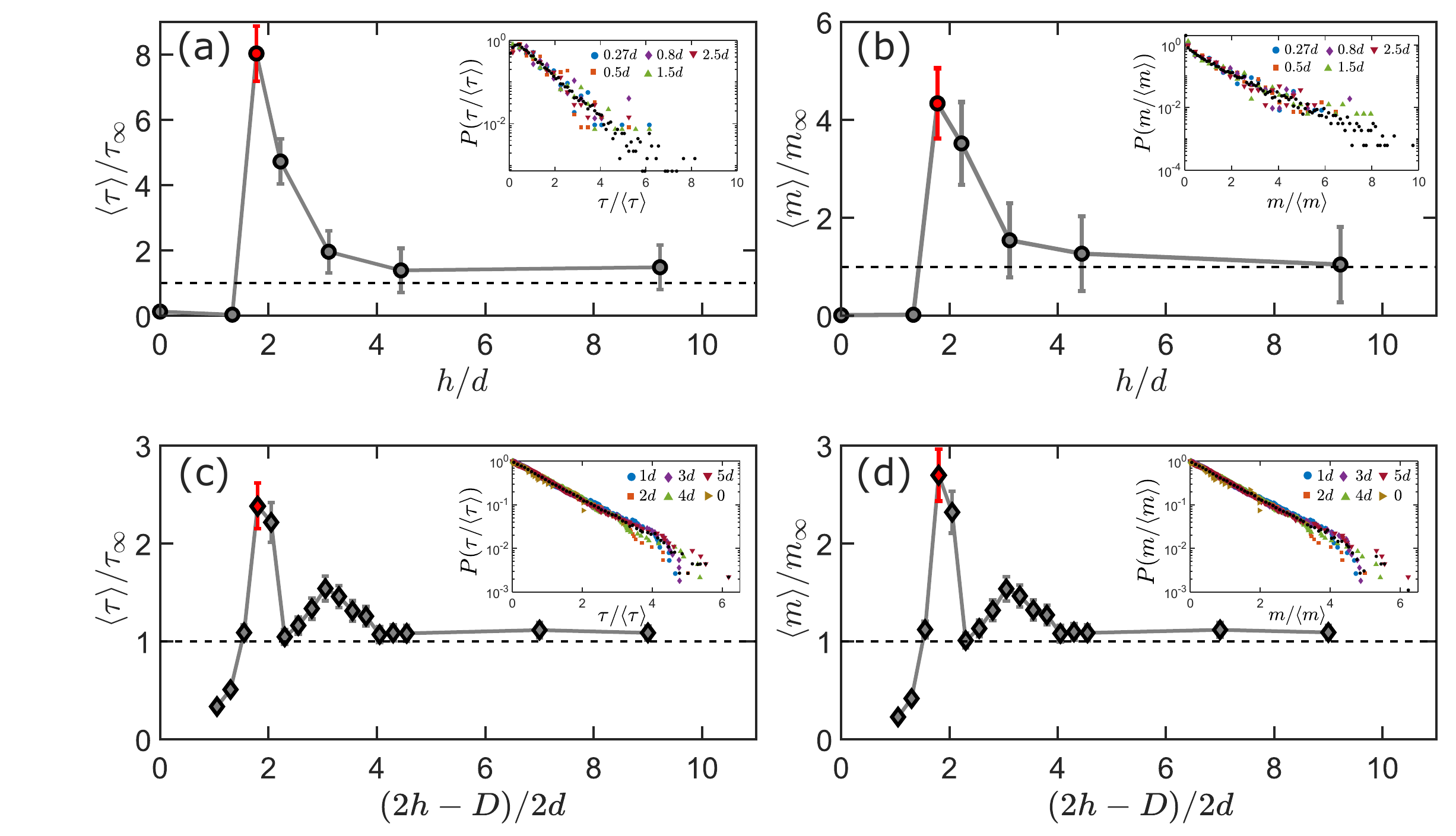}
    \caption{\textbf{Flow as a function of obstacle position} (a) and (b) show experimental data for the average duration ($\langle \tau \rangle /\tau_{\infty}$) and average mass ($\langle m \rangle/m_{\infty}$) in one discharge, both plotted as a function of tip height $h$ of the obstacle in units of particle diameter $d$). In (c) and (d) we plot the corresponding data for the simulation. 
    Here, $\tau_{\infty}$ and $m_{\infty}$ represent the duration of the flow and the discharge from the silo without obstacle, respectively. The obstacle diameter is held fixed at $D=2d$ and $D=0.5d$ for simulations and experiments respectively, and the outlet diameter in both cases is $a=3d$. The red point indicates the height $h_{pk}$ where the suppression of clogging is maximized.
    In the inset, we show probability distributions $P(m/m_{\infty})$ and $P(\tau/\tau_{\infty})$ of the mass and duration of the discharges.  To improve statistics, data from different rod diameters and heights are combined in the inset, where the black points represent the average of the different diameters.
    }
        \label{fig:Fig2}
\end{figure}

\textbf{Measurements}  In both experiment and simulation 
we gather data on the statistics of the duration $\tau$ of the discharge before the flow clogs, as well as the discharge size, as measured by the total mass $m$ of particles that leave the hopper between clogging events.  The duration of the discharge is monitored using an audio microphone, an example of whose output is shown in Fig. $\ref{fig:Fig1}$ (c). The microphone is positioned to pick up a signal from the grain collisions alone. To measure the mass $m$ that is discharged, an electronic weighing balance is placed beneath the outlet, and the cumulative deposited mass is read out just prior to a hammer strike. The discharged mass is measured with an accuracy of $\pm 0.1$ gram corresponding to a resolution of about $10$  particles; in the simulations, of course, single particles are resolved.


\begin{figure}
    \includegraphics[width=.95\textwidth]{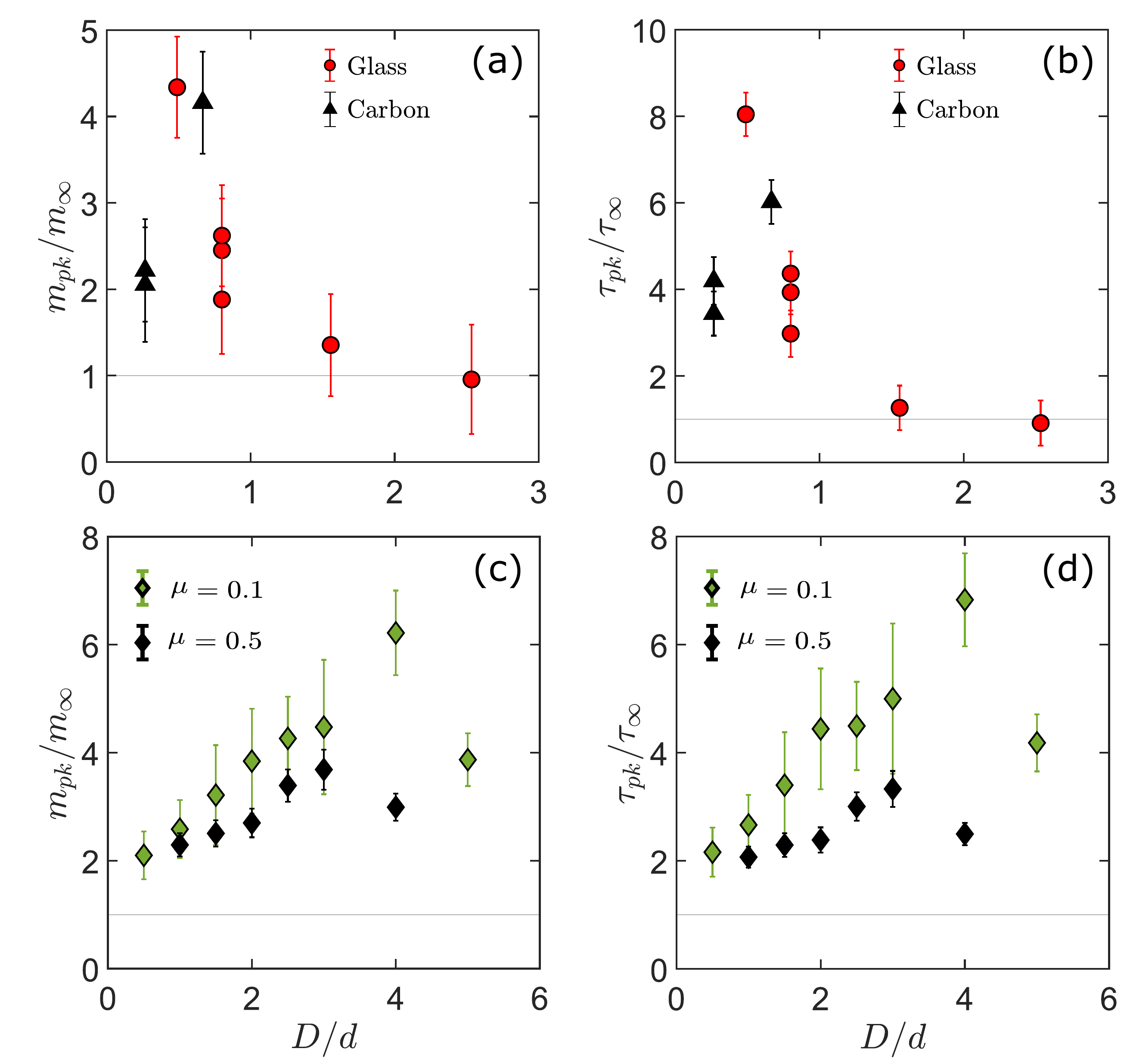}
      \caption{\textbf{Clog suppression as a function of obstacle size} Normalized peak duration ($\tau_{pk}/\tau_{\infty}$)  and peak discharge ($m_{pk}/m_{\infty}$) are plotted against obstacle diameter $D$ (in particle diameters $d$) for (a, b) Experiment and (c, d) Simulation. Labels in (a, b) represent different rod materials, while those in (c, d) indicate varying inter-particle friction coefficients $\mu$.}
        \label{fig:Fig3}
\end{figure}


As a first characterization of the effect of the obstacle on the flow, we plot both the mass, $m$, and the duration, $\tau$, averaged over a few hundred clogging events at several different positions of the obstacle along the vertical axis.  As shown in Fig. \ref{fig:Fig2}, these quantities vary strongly with the height $h$ of the obstacle above the opening in both simulation and experiment ($h$ is defined from the center of the sphere, and from the tip of the rod, respectively).  At small $h$, the flow is strongly suppressed by the obstacle, which plugs the opening.  As expected, when the obstacle is far above the opening ($h \rightarrow\infty$), both the discharged mass and duration revert to the values $m_{\infty}$ and $\tau_{\infty}$ obtained in the absence of an obstacle.  However, at intermediate values of $h=h_{pk}$, there is a peak in both $m$ and $\tau$, showing that the suppression of clogging can be optimized by obstacle placement, and demonstrating that in a fully 3-dimensional silo discharge an appropriately placed object -- either cylindrical or spherical -- facilitates flow, just as in the heavily-studied 2D geometry. 

The effect is clear in both experiment and simulation, though the amplitude of clogging separation, namely the value of $m_{pk}/m_{\infty}$ and $\tau_{pk}/\tau_{\infty}$ is different between them.  For the parameters shown in Fig.\ref{fig:Fig2} (a), a rod-tip at the optimal position ($h=h_{pk}$) produces nearly an order of magnitude enhancement in the duration of a clog, even though the projected area of the obstacle is only about 2\% of the area of the outlet. The simulation data also consistently shows a secondary peak of lower magnitude at $h>h_{pk}$. We return to this feature later as it reinforces our picture of the mechanism of clog suppression. 

The enhancement in the mean discharge mass and time does not come from unusually large fluctuations.  To the contrary, as shown in the insets to Fig.\ref{fig:Fig2},  their probability distributions $P(m)$ and $P(\tau)$ remain exponential at all values of $h$, just as they are in the absence of an obstacle\cite{clement2000jamming}. This suggests that the unclogging is statistically a Poisson process where each discharge is independent of the previous clogging event.  The effect of the obstacle is thus to enhance the mean, and the fluctuations merely scale accordingly.  To a first approximation, the mean flow rate during a discharge, $m/\tau$, is only modestly altered by the presence (and location) of the obstacle, as has also been observed in 2D (see Supplementary Material Figure \ref{suppl:fig:suppl_Flow_rate}).  


\begin{figure}
    \includegraphics[width=.95\textwidth]{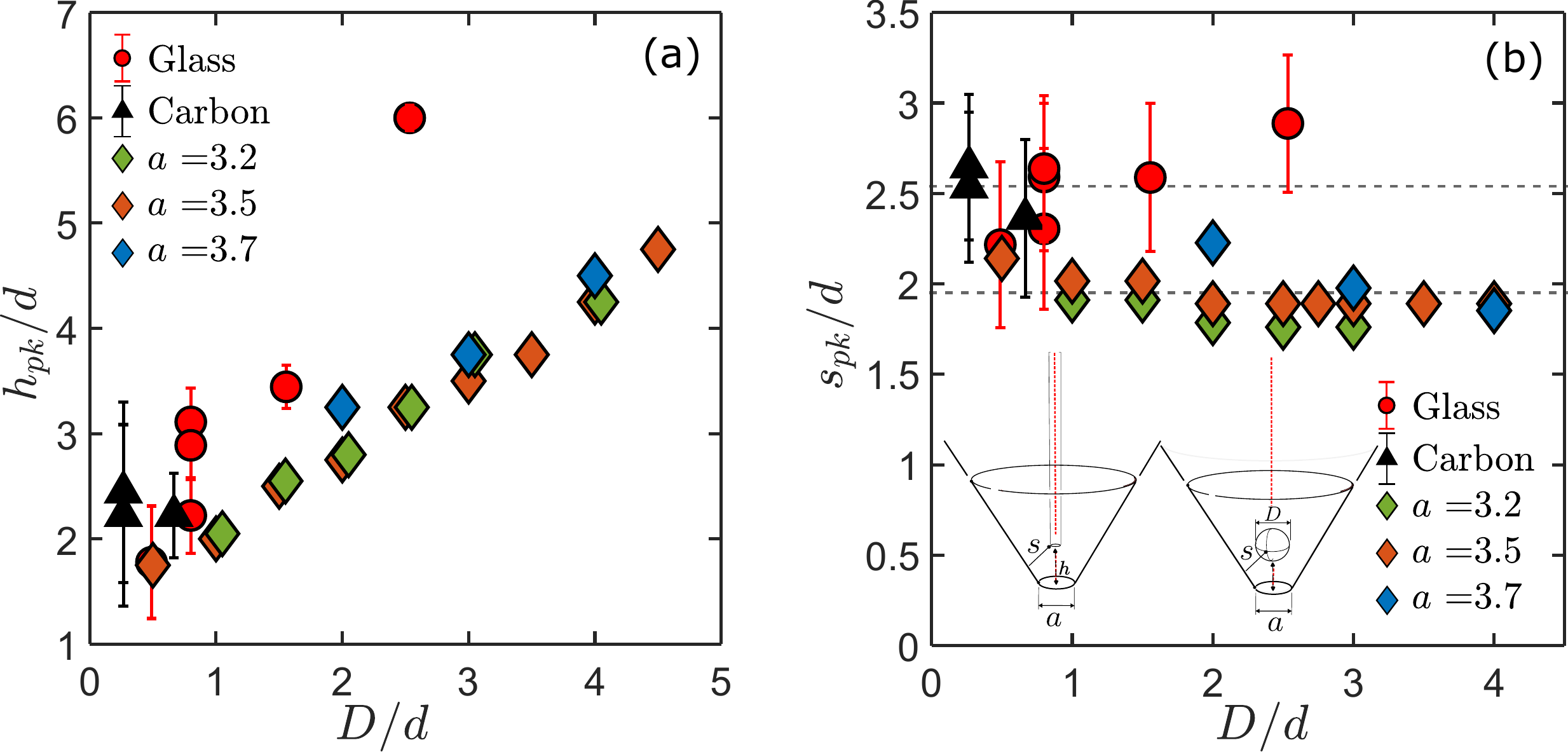}
    \caption{\textbf{Optimal location preserves shortest distance to wall} Variation of  (a) the peak height $h_{pk}$  and (b) the shortest distance $s_{pk}$  from the surface of the obstacle to the inner surface of the hopper as a function of  $D/d$. Experimental data are shown as circles and triangles, representing different rod materials with a hopper opening of $a = 3.2d$. Diamonds denote simulation data with varying opening sizes $a$. }
        \label{fig:Fig4}
\end{figure}


\begin{figure}
    \includegraphics[width=.95\textwidth]{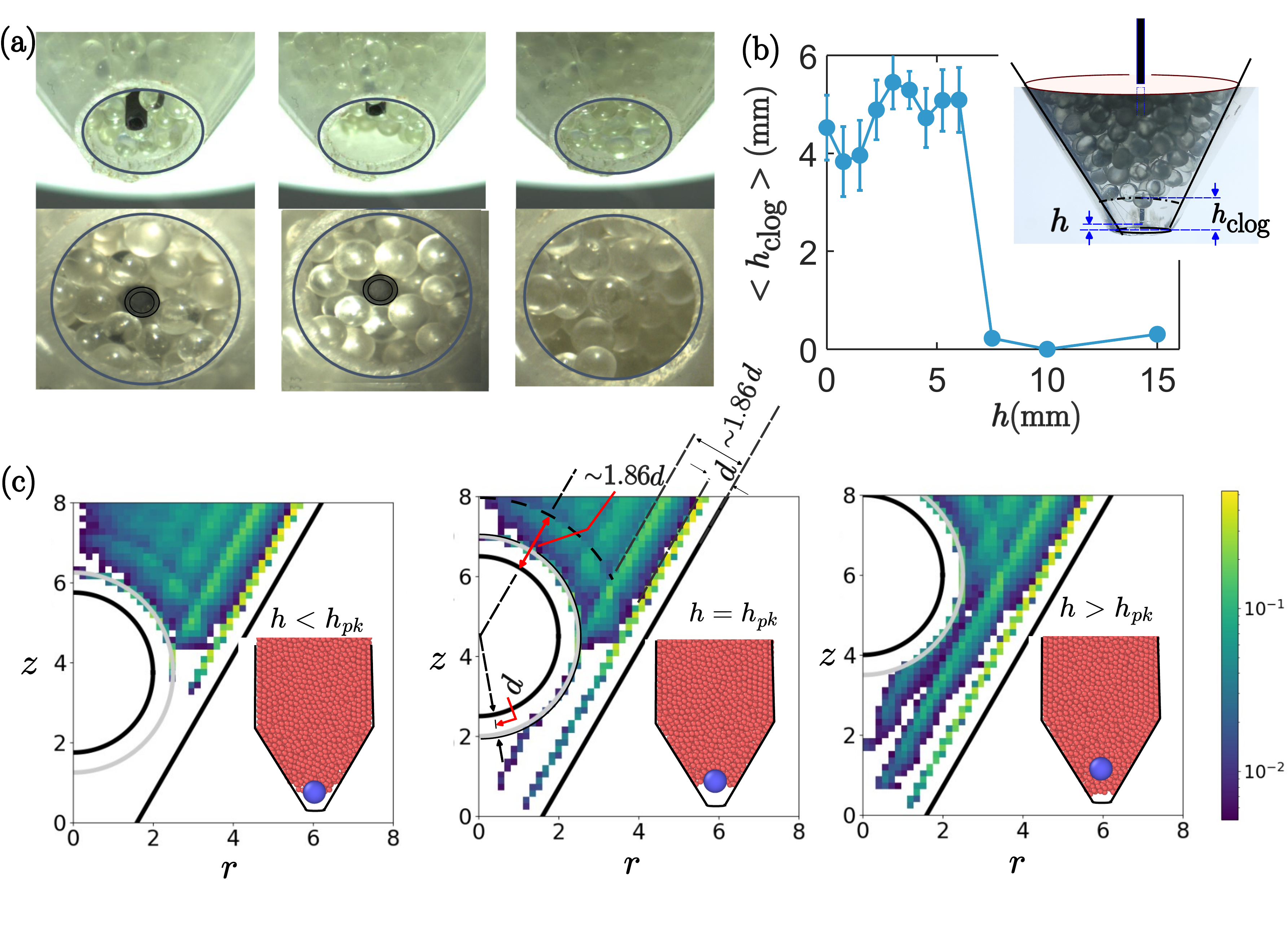}
    \caption{ \textbf{Optimal clogging suppression is set by geometrical considerations} Experiments:  (a)  The top and bottom panels show representative images of the jammed configurations of the particles for $h<h_{pk}$,  $h\approx h_{pk}$ and $h>h_{pk}$ as viewed from two different camera positions. (b) Variation of the average height of the clog $\langle h_{clog}\rangle$ as measured from the plane of the orifice as a function of height $h$ of the obstacle.   The inset shows a representative image of the clog near the orifice.  \textbf{Simulations:} (c) Density of grains in the hopper in the clogged state as a function of r (distance to the central axis) and z (height above the opening). Here the grains clog with the obstacle for \(h<h_{peak}\), without the obstacle for \(h>h_{peak}\), and both with and without the obstacle for \(h=h_{peak}\).}
        \label{fig:Fig5}
\end{figure}

Just as the location of the obstacle affects the magnitude of the effect, so too does the size of the obstacle, as we show in Fig.\ref{fig:Fig3}.  In the top row, we show experimental data for the enhancement in mass and duration, $m_{pk}/m_{\infty}$ and $\tau_{pk}/\tau_{\infty}$, as a function of rod diameter $D$.  
There is a sharply defined peak at $D/d \approx 0.6d$, that is, the effect of the obstacle is maximized by a rod whose diameter is even less than that of a grain. The simulation counterpart, in the bottom row of Fig.\ref{fig:Fig3}, shows a similar effect for a spherical obstacle for two values of friction, $\mu = 0.1$ and $0.5$.  The effect is less sharply peaked, but the optimal size of the sphere does not change much over this range of friction coefficients, which spans most commonly used laboratory granular materials. The magnitude of the effect is larger for the lower friction coefficient ($\mu = 0.1$). To get an estimate of the experimental friction coefficient, we pull a rod out of a granular packing while varying the normal load  (see  Supplementary Material  Section \ref{suppl:sec:friction}  and Figure 
\ref{suppl:fig:Fricrion}).  Even at large normal loads, we measure relatively small friction coefficients $\mu \sim 0.01$, suggesting that the spheres predominantly roll on the rod.

Having established the robustness of the clog suppression in 3D over two distinct obstacle geometries and different material properties, we turn to the question of what dictates the optimal location of the obstacle.  In Fig.\ref{fig:Fig4}(a) we show that the optimal vertical location $h_{pk}$ varies substantially as object size is altered.

However, in Fig.\ref{fig:Fig4}(b) we replot the location in terms of  $s_{pk}$, the shortest distance between the obstacle and silo wall at the optimal placement.  The optimal $s_{pk}$ is nearly constant as the obstacle size $D$ is changed substantially; for all values of $D$, $s_{pk} \approx 2.4$ in the experiments, and $\approx 1.9 \pm 0.15$ in simulations.  This nearly invariant value of $s_{pk}$ suggests that there is an underlying geometrical explanation for the optimal placement for clog suppression. 

We pursue this idea by presenting evidence that the optimal location divides two regimes of clogging: for $h>h_{pk}$, the clog forms beneath the obstacle, and the obstacle does not form part of the arch (or other spanning structure) and for $h<h_{pk}$, the clog forms around the obstacle. For $h \approx h_{pk}$, the optimal position, the clog forms just at the tip of the rod.  Images exemplifying these three behaviors from the experiment are shown in Fig. \ref{fig:Fig5}(a). 
To better quantify this claim, we show in Fig. \ref{fig:Fig5} (b) a method of locating the arch by side view images of the funnel. While we cannot obtain the 3D spatial structure of the arch in the experiment, we can from the 2D planar projection infer the height of the arch $h_{clog}$ as shown in Fig. \ref{fig:Fig5} (a) (further details are available in Supplementary Material Fig.\ref{suppl:fig:suppl_schematic_fig5_new},
\ref{suppl:fig:suupl_all_theta_signal}).  The outcome can be averaged over a statistically large number of clogs and a mean arch height identified for all values of $h$.  In Fig. \ref{fig:Fig5} (b), we can see that the arch location drops sharply to the opening as $h$ crosses $h_{pk}$.  The evidence thus implies that at $h=h_{pk}$ the most likely location of the arch is in the most precarious spot: any slippage beyond the tip of the rod, and the arch loses the support of the obstacle. 

In Fig \ref{fig:Fig5}(c), we show parallel evidence from the simulation via density plots that average over a large number of clogs at $h$ less than, approximately equal to, and greater than $h_{pk}$, as well as a cross-section through an example of a clog at each of these obstacle locations.
Once again, these figures show that the optimal location, $h=h_{pk}$, marks a transition in the location of the clog from being above the equator of the spherical obstacle, to below it. The simulation data reveals additional information about the 3-dimensional structure of the clog: there is strong alignment in layers near the wall (Supplementary Fig. \ref{suppl:fig:suppl_t_dist}), and also layering along the spherical surface of the obstacle. At the optimal height, the clearance between the side-wall and the obstacle $s=s_{pk}\approx 1.9$ just allows a two-particle wide clog to form where one particle rests against the obstacle and another against the hopper
surface.  Any slippage below this location leads to a widening of the space available, and thus the clog loses support against the wall. Furthermore, it is possible to see a third, more diffuse layer, leading to a three-particle-wide clog at which likely is the origin of the second peak (corresponding to \( s_{pk} \approx 2.5 \) in the simulation data in the lower row of Fig. \ref{fig:Fig2}. A simple 2D model of the packing of disks clarifying this geometric reasoning is given in the Supplementary Material (Fig. \ref{suppl:fig:suppl_model}). 

These observations on layering against the wall and against the obstacle also lead to a natural explanation for why the second peak is absent in the experimental data shown in Fig. \ref{fig:Fig2} and why the optimal $s_{pk}$ is numerically slightly different in experiment and simulation ($2.45$ rather than $1.9$).  The grains in the simulation are monodisperse spheres, whereas the experimental grains range in diameter from $2-2.5 mm$, and are not perfectly spherical. We surmise that polydispersity suppresses the formation of well-defined layers. The consistent value of $s_{pk}$ within each system reinforces a geometric picture where the most probable arch in the available space \cite{to2001jamming} is set only by the shape and size of the grain. 
 

Finally, the geometric picture pinpoints a sufficient ingredient for clog suppression to work: the obstacle creates a diverging flow geometry at some point (the tip of the rod, the equator of the sphere) near the outlet. The effect is optimized when the most probable clog location is at this vulnerable spot.  This picture does not require some elements that have previously been identified as the key mechanism for clog-suppression; for example, in the rod geometry, there is no enhancement of the granular temperature by scattering in the direction transverse to the flow, the flow does not slow down anywhere, and there is no discernible `waiting room'. These factors can possibly be tuned to enhance the effect, but they are not a necessary ingredient for clog suppression. \cite{escobar2003architectural,alonso2012bottlenecks,
gella2022dual,
gao2019understanding}

We have thus established that the beautiful and surprising effect of clog-suppression by an obstacle carries over into a 3-dimensional system, opening the way to practical applications in the delivery of granular matter in a controlled manner even close to a putative clogging transition. The effect is robust: between experiment and simulation, we have explored the role of friction, polydispersity of grains, and shape of the obstacle.  We have also found (see Supplementary Material) that the magnitude of clog-suppression is not strongly affected by the rigidity of the rod and the suspension mechanism. The idea that clogging is inherently a geometric effect \cite{to2001jamming} and is not fundamentally a dynamical phenomenon has been explored for example in work that shows that inertia \cite{gao2021enhanced, koivisto2017effect, gella2022dual} is not a dominant factor in clogging.  Similarly, the role of outlet shape and corner geometry \cite{gao2019understanding, hanlan2024cornerstones} has also been explored. A very practical advantage our work reveals is that even a slender obstacle can play a significant role. Thus an obstacle can be inserted into a hopper when necessary, without interruption of a flow process.  The underlying geometric idea of introducing any locally divergent flow geometry also points to the generality of the mechanism, independent of particle interactions, inertia, or a background solvent; this greatly broadens the potential applicability of intervening with an obstacle in other particulate flows \textit{e.g.} suspensions or foams. 

\appendix*
\section{Methods}

\section{Simulation}
     In the simulation, the particle-particle and particle-wall (both boundary wall and obstacle) interactions are governed by Hertzian contact interactions \cite{DEM} with viscous damping in both normal and tangential directions \cite{silbert2001}. The interparticle force is:
\begin{equation}    
\textbf{F}_{hz}=\sqrt{d\delta/4}[k_n\delta \boldsymbol{n}_{ij}-m_\textrm{eff}\gamma_n \boldsymbol{v}_n-k_t \boldsymbol{\Delta s_t}-m_\textrm{eff}\gamma_t \boldsymbol{v}_t]
\end{equation}
Here $\delta$ is the overlap between contacting particles $i$ and $j$,  $\boldsymbol{n}_{ij}$ the unit vector along the line connecting the two particles, and $\boldsymbol{v}_n$ and $\boldsymbol{v}_t$ the normal and tangential components of their relative velocity. A similar equation holds for particle-wall interactions. 

The normal and tangential spring constants used in our simulations are
$k_n=7 \times 10^{4} N/m$,
$k_t=9 \times 10^{4} N/m$, with normal and tangential damping
$\gamma_n=8 \times 10^{5} s^{-1}$ and 
$\gamma_t=\gamma_n/2$. The tangential
displacement $\boldsymbol{\Delta s_t}$ is measured from the point where contact is originally made.  The contact slips when the ratio of tangential to normal contact forces exceeds the coefficient of static friction $\mu$ and $\boldsymbol{\Delta s_t}$ is reset to zero. 
    The system has periodic boundary conditions in the z direction:  grains that leave the hopper are reintroduced at the top to ensure a constant number of grains in the system. All lengths are measured in units of the particle diameter, set to 1, and the simulation timestep is $3.2\times10^{-6}s$.  Aside from the position and velocity of all particles, we also determine measure the contact forces between particles, and the obstacle and side walls when the hopper is clogged.

\section{Experiment}
\paragraph{Experimental setup:}  The cylindrical section of the hopper has a diameter of $W = 94 \, \text{mm}$ and a length of $L = 500 \, \text{mm}$. The hopper is initially filled to capacity, and the experiment is conducted until the grain surface drops to 300 mm from the top. At this point, the experiment is paused, the hopper is refilled, and then the experiment resumes.  
The solenoid (connected to an Arduino), microphone, and weighing balance are interfaced with a computer. A Matlab script automates hammer strike timing and the data acquisition process. The interval between successive hammer strikes is approximately 20 seconds. The   duration of the discharge is monitored using audio signal from the microphone (Boya ByM1 Auxiliary Omnidirectional Lavalier Condenser Microphone)  sampled at 44kHz.  The  upper envelope of the denoised signal is thresholded to produce a binary signal which is a train of rectangular pulses. Avalanche duration is determined by the width of the corresponding pulse in the binary signal (see Fig.\ref{suppl:fig:audio_signal}).

 \paragraph{Clamping conditions:} Various clamping conditions for the rod were tested to optimize flow control  (see Figs.\ref{suppl:fig:suppl_fig_clamping},\ref{suppl:fig:boundary_conditions} of the SI). For the main body of this study, we use a clamping condition in which the upper tip of the rod is attached to an inextensible string, allowing it to behave like a pendulum. This setup enables the rod to dynamically adjust its position as the granular discharge proceeds, potentially affecting the flow rate and pattern. However, in practice, we found that the pressure of the surrounding flow centered the rod in the hopper and that the rod stayed stationary along the axis to within measurement precision.

 \section{Acknowledgments}  NM acknowledges funding through NSF-DMR 2319881. SG acknowledges the support of the Department of Atomic Energy, Government of India, under Project No. 12-R \&DTFR-5.10-0100. We thank Anit Sane for technical support.

\bibliographystyle{apsrev4-1}
\bibliography{hopper}
\end{document}


\title{Supplementary Information : Facilitating a 3D granular flow with an obstruction}
 \author{Abhijit Sinha, Jackson Diodati, Narayanan Menon, Shubha Tewari, Shankar Ghosh}
\affiliation{}%

\date{\today}

\begin{abstract}

\end{abstract}
\maketitle

\renewcommand{\thefigure}{SF\arabic{figure}}
\renewcommand{\thesection}{S\arabic{section}}
\renewcommand{\thesubsection}{SS\arabic{subsection}}

\renewcommand{\theequation}{S.\arabic{equation}}

\section{Friction measurements}\label{suppl:sec:friction}
\begin{figure}[h]

   \includegraphics[width=.7\textwidth]{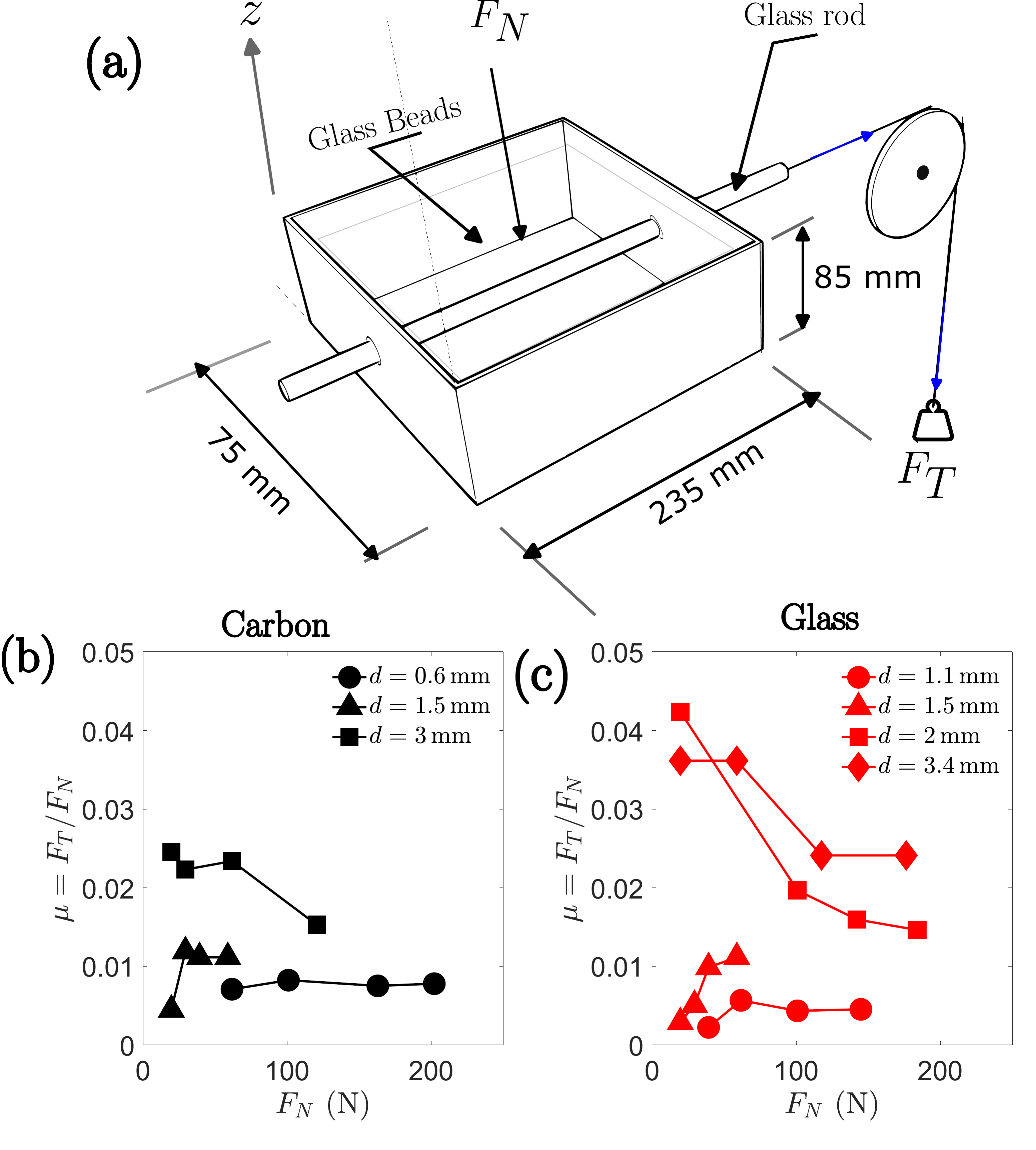}
    \caption{\textbf{Friction Measurement:}(a) Setup of our experiment.  The figure shows the variation of the friction coefficient between rod and glass beads as a function of the normal force $F_N$ for different rod diameters $D$. The panel (b)  is for carbon rods and (c) is for glass rods.}
    \label{suppl:fig:Fricrion}
\end{figure}

To measure the friction coefficient of the rod with respect to the glass beads, we filled a rectangular container (dimensions: $235 \times 75 \times 85 mm$) with glass beads (diameter $d=2-2.5 mm$).  Holes were drilled on the opposite faces of the container to allow a rod (of glass or carbon) to pass through the two opposite walls of the container smoothly. One end of the rod was connected to a string that ran over a pulley. 
A flat, rigid plate was placed on top of the glass beads, and weights were added to this plate to vary the normal force $ F_N $.   The tension in the string was increased by attaching weights to the free end of the string until the rod began to slide at a tension $ F_T$. This experimental setup is schematically shown in Fig. \ref{suppl:fig:Fricrion}a.  The coefficient of static friction $ \mu $  between the rod and the particles was computed from the 
$
\mu = F_T/F_N
$. Figure \ref{suppl:fig:Fricrion}b shows the relationship between the measured coefficient of static friction, $\mu$, and the rod diameter, $D$, for two materials: glass and carbon.  We make a few important observations about the data: 
(1)  Carbon rod shows a smaller frictional resistance as compared to the glass rods. 
(2) For both rods and for all rod diameters,  $\mu$ appears to reach an asymptotic value at large values of $F_N$.
(3) The friction coefficient increases with the rod diameter $D$.  This is consistent with an increase in the frictional resistance as more particles make contact with the rod. 
(4) The measured friction coefficients are exceptionally low ($\mu < 0.05$)  which leads us to infer that the frictional yielding is from rolling, rather than sliding, contacts. 

\begin{figure}[htb]
    \centering
    \includegraphics[width=.75\textwidth]{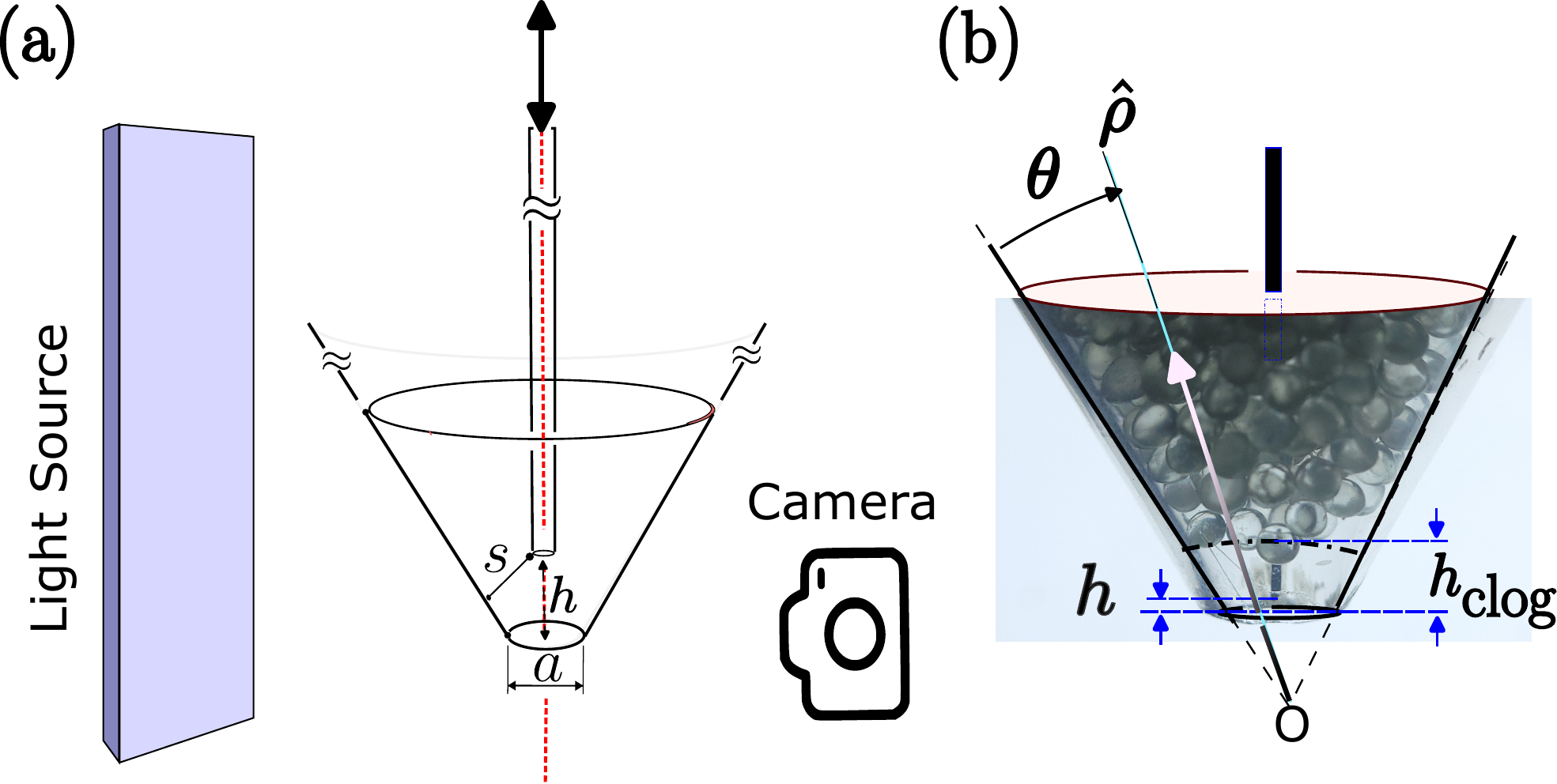}
    \caption{ (a)Schematic of the experimental setup to calculate arch height distribution (b) The closeup of the clog near the orifice  along with the geometric descriptors that are used to compute $h_{\textrm{clog}}.$}
       \label{suppl:fig:suppl_schematic_fig5_new}
     \end{figure}

\section{Imaging clogged arches}


To image the structure of the clog we used a modified apparatus: The rod was attached to a motorized translation stage and placed at a height $h$ above the outlet. It then periodically moves up (at speed $5mm/sec$) by a distance $\delta h  \pm \sim 1$ mm, waits for a time period $\tau_{r}$, and then moves down again.  The up-and-down motion provides a mechanical perturbation to break clogs.  Before moving down from the higher position, a side-view image is taken of the region near the outlet.

The resting period is chosen such that the flow is clogged when the image is taken, i.e., $\tau_{r} \gg \langle \tau \rangle$, where $\langle \tau \rangle$ is the mean time during which the grains flow.  For every value of $h$, $100-250$ images (resolution-$1280\times960)$ are taken with a USB-Teledyne Lumenera camera. The entire setup is automated using Matlab.  
 A schematic of the experimental setup, including the lighting and the camera position is shown in Fig. \ref{suppl:fig:suppl_schematic_fig5_new}. 
\begin{figure}[t]
 \includegraphics[width=.7\textwidth]{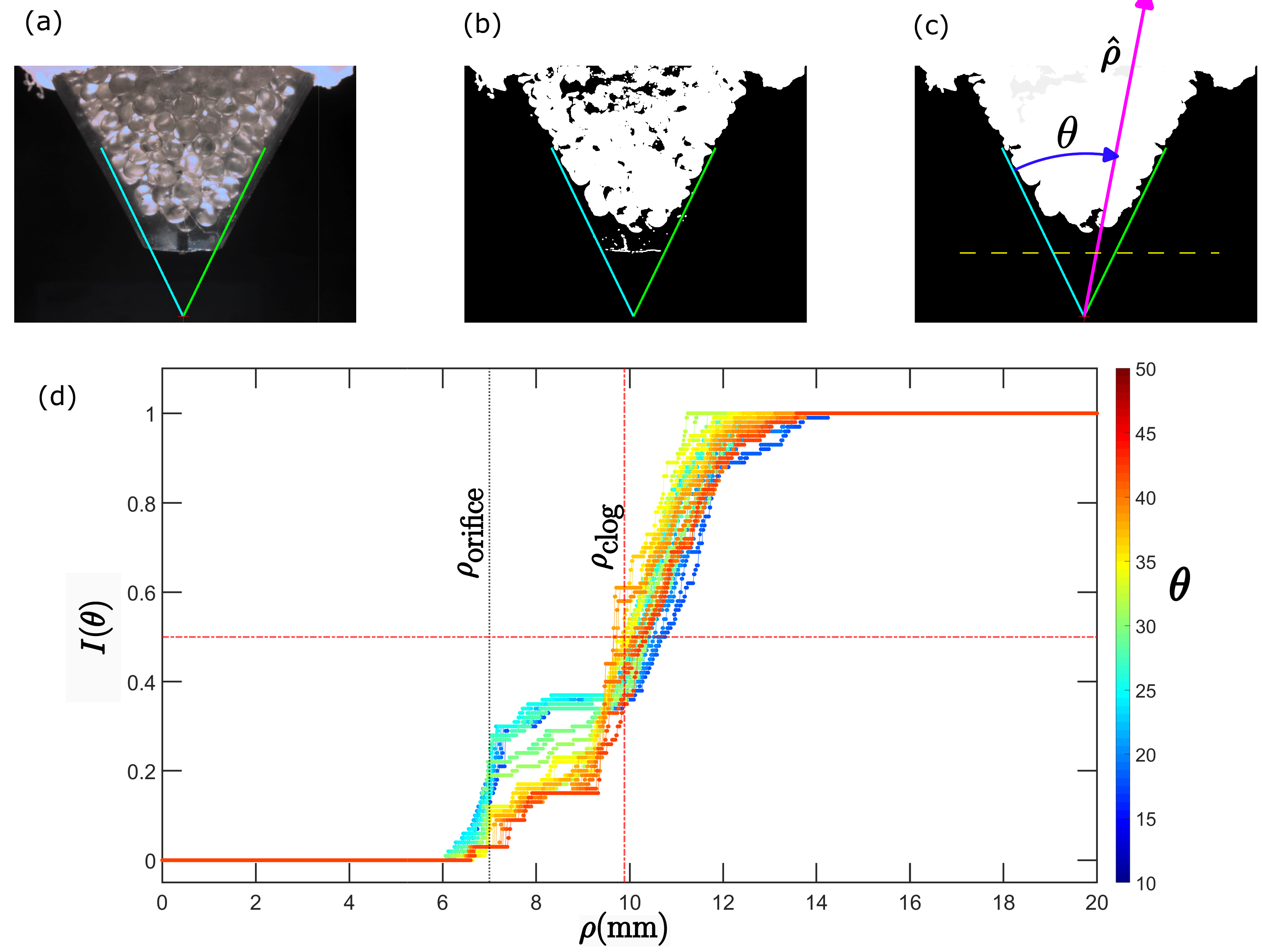}
   \caption{\textbf{Image Analysis} (a) Raw image. (b) Thresholded and binarized image. (c) Image after filling up the holes in (b) and choosing only the maximum cluster of ones. (d) $I(\theta)$ is the average intensity along the ray $(\theta)$ (different colors indicate different theta) vs radius $\rho$. Here $\theta$ is taken to the left end of the Hooper. The vertical lines at $\rho_{\textrm{orifice}}$ and $\rho_{\textrm{clog}}$  denote the values of $\rho$ at which it meets the plane of the opening of the orifice and the boundary of the jammed interface. Thus $h_{\textrm{clog}}=\rho_{\textrm{clog}}-\rho_{\textrm{orifice}}.$}
    \label{suppl:fig:suupl_all_theta_signal}
  \end{figure}   
  
 \subsection{Image analysis for detection of the jamming interface}  
 Figure  \ref{suppl:fig:suupl_all_theta_signal}  shows the process involved in determining the height $h_{clog}$ of the jamming interface. The input image $I$ is first converted to grayscale. Next, a mean filter with a neighborhood size of 5 is applied to the grayscale image to smooth it. The filtered image is then thresholded, producing a binary image. Interior holes in this binary image are filled. Then, the connected components of the binary image are identified and their sizes calculated. All pixel values belonging to connected components, except for the largest one, are set to zero, effectively isolating the largest connected component.
In the next step, the algorithm loops over a range of angles $\theta$, and for each $\theta$, it iterates over a range of radial $\rho$ distances, thus producing a set of diverging rays $\hat{\rho}(\theta)$ from the origin. For each combination of angle and radius, a ray is defined.  The algorithm records the radial positions and pixel values $I(\rho, \theta)$ along the corresponding ray $\hat{\rho}(\theta)$. The ray $\hat{\rho}(\theta)$  meets the plane of the opening of the orifice and the boundary of the jammed interface at $\rho_{\textrm{orifice}}$ and $\rho_{\textrm{clog}}$, respectively.
The average position of the jammed interface is determined by the condition:
\[
I(\rho_{\textrm{clog}}) = \langle I(\rho, \theta) \rangle_\theta = 0.5,
\]
whereas the onset of the curve's knee in $\langle I(\rho, \theta) \rangle_\theta$ corresponds to the plane of the hopper's outlet.  The vertical lines in Fig. \ref{suppl:fig:suupl_all_theta_signal} show the positions of  $\rho_{\textrm{orifice}}$ and $\rho_{\textrm{clog}}$.  The height of the clog is measured from the orifice plane, thus 
\[h_{\textrm{clog}}=\rho_{\textrm{clog}}-\rho_{\textrm{orifice}}.\]

 Figure 
\ref{suppl:fig:suppl_arch_hist} shows the height distribution $p(h_{clog})$ of the arch for various values of obstacle diameter $D$ and height $h$. For $h < h_{pk}$, clogging predominantly occurs near the tip of the obstacle, with $p(h_{clog})$ peaking around $h_{clog} \sim h$. However,  When $h \gg h_{pk}$, the clogging position descends to the region near the orifice, with $p(h_{clog})$ peaking at approximately $h_{clog} \approx 0$. These trends are consistent across all values of $D$.

\subsection{Audio signal processing}\label{suppl:Audio}

The duration of granular discharge is monitored using an audio microphone (Boya ByM1 Auxiliary Omnidirectional Lavalier Condenser Microphone).
The signal $ S_1 $ is recorded by the microphone at a sampling rate of 44 kHz
A moving filter $ F $ is applied to $ S_1 $ to reduce noise, producing the filtered signal $ F(S_1) $. The upper envelope of the absolute value of the filtered signal is then computed as follows:

\[
S_2 = \text{Env}_\text{upper}\left( |F(S_1)| \right)
\]

\begin{figure}
  \includegraphics[width=.7\textwidth]{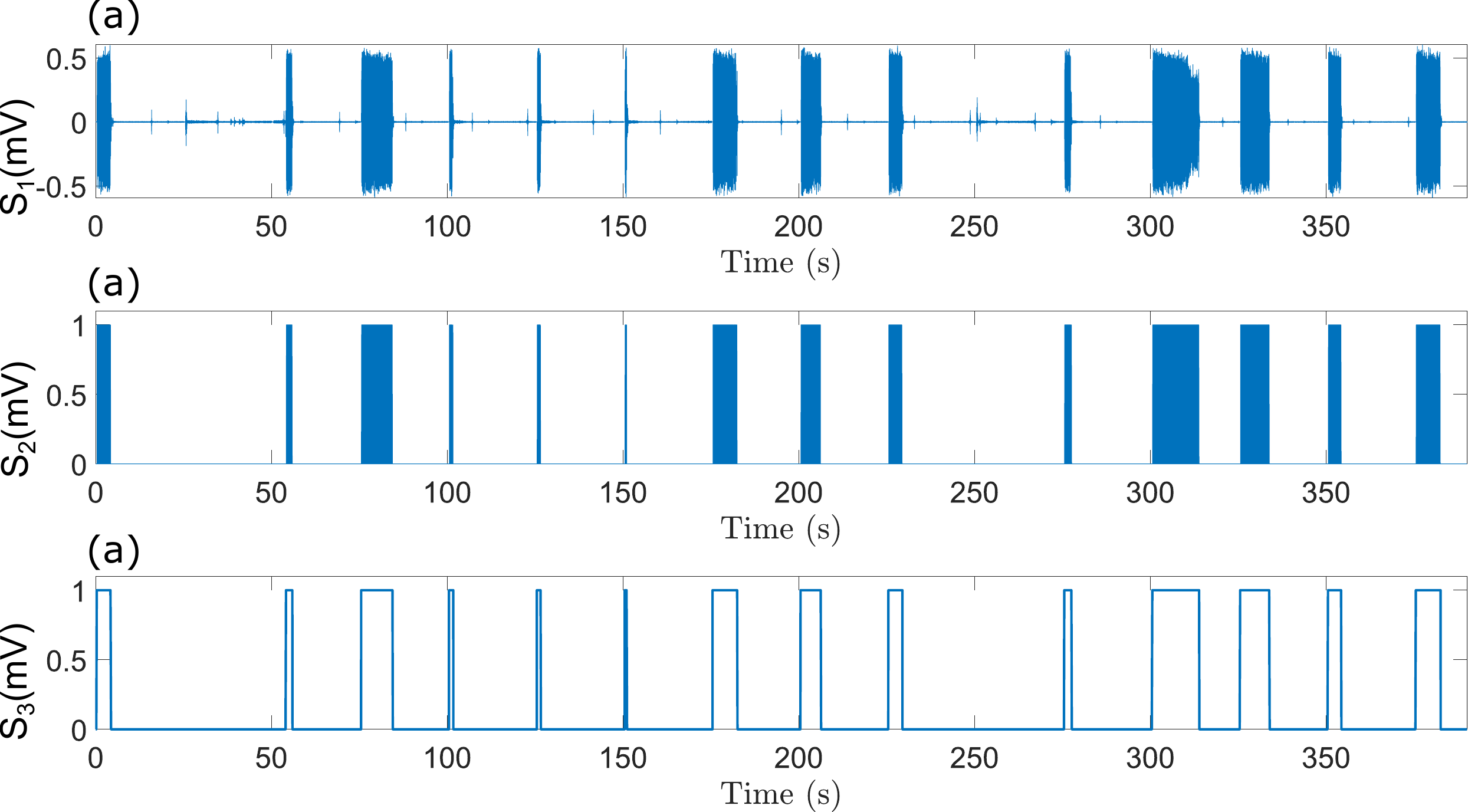}
   \caption{\textbf{Audio Signal}.Here (a) S1 is the audio signal vs Time,(b) S2  is the filtered signal vs Time, and (c) S3 is the envelope of the signal vs Time. }
    \label{suppl:fig:audio_signal}
  \end{figure}

The signal $ S_2 $ is subsequently thresholded (see Fig. \ref{suppl:fig:audio_signal}) to generate the binary signal $ S_{th} $, which consists of a train of rectangular pulses. 
The duration of an avalanche is determined by the width of the corresponding pulse in $ S_{th} $.

\section{Additional results from simulation and experiment}
\begin{figure}[ht]
    \centering
    \includegraphics[width=.9\textwidth]{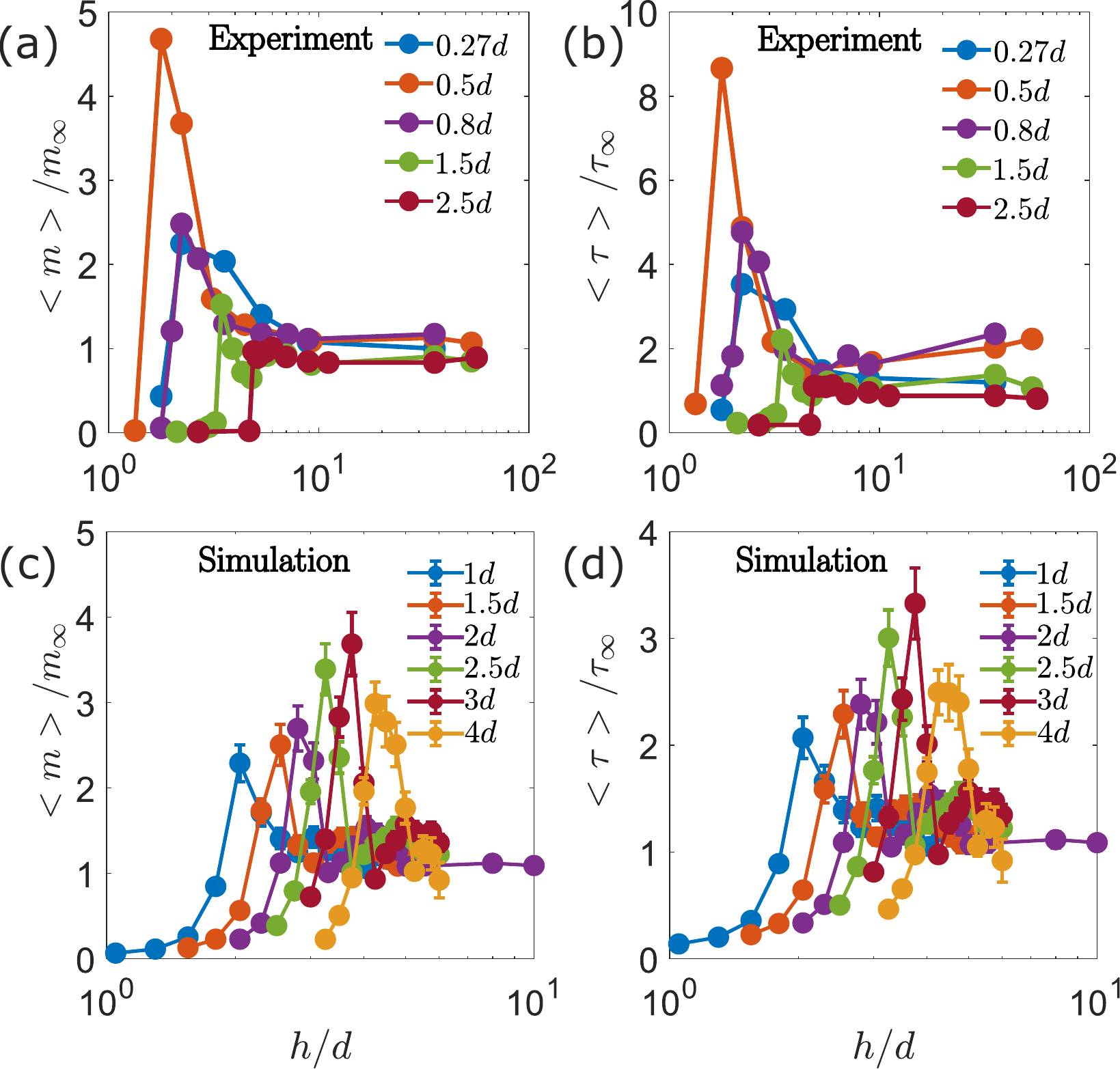}
    \caption{\textbf{Average Discharge vs Height and  Average Duration vs Height:} The variation in average mass flow (panels (a) for experiment and (c) for simulation) and the flow duration $\tau$ (panels (b) for experiment and (d) for simulation) following the mechanical impulse is presented as a function of the height $h$ of the lowest point of the obstacle. }
            \label{suppl:fig:suppl_mass_time}
    \end{figure}

 In  Fig. \ref{suppl:fig:suppl_mass_time}  we present the data corresponding to the variation of the discharge mass and duration for obstacles of varying diameters and heights at which they are placed. The data from both experiments and simulations show that for any given diameter of the obstacle, there is an optimal positioning of the obstacle at which the discharge is the maximum.  The corresponding value of $h$ is referred to as  $h_{pk}$. The discharge mass and the discharge duration at $h=h_{pk}$ are plotted in Fig.3 of the main paper. Moreover, there also exists a most effective diameter of the obstacle for which the effect is maximum.   The variation of $h=h_{pk}$ as a function of the obstacle diameter is plotted in Fig.4 of the main manuscript.

\begin{figure}
    \centering
    \includegraphics[width=0.5\linewidth]{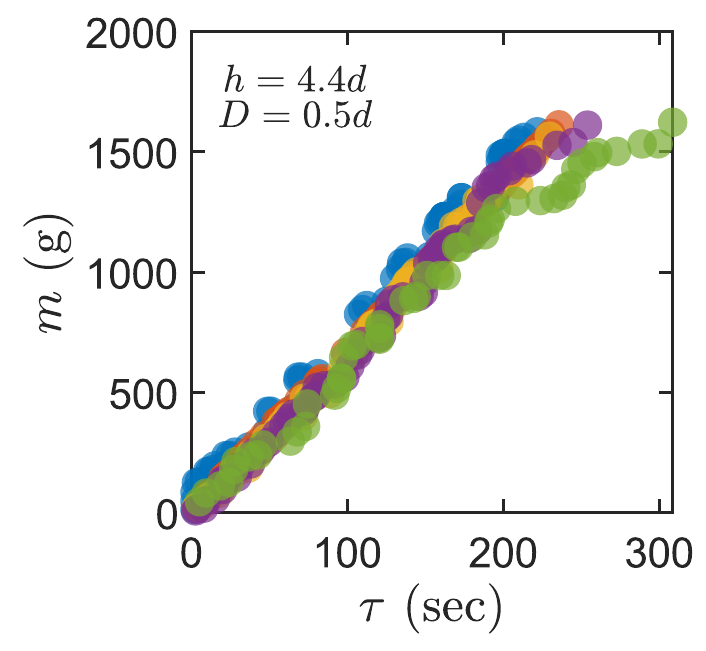}
    \caption{\textbf{Determining the Flow Rate  :} The individual instances of  discharged mass $m$ vs. the discharge duration $\tau$  shows a linear trend. The  colors correspond to different   experimental  realization with the same parameters, $h=4.4d$ and $D=0.5d$. The  flow rate $\dot{Q}$ is determined by the slope of the straight line fitted to this data. }
    \label{suppl_mass_vs_time_obstacle}
\end{figure}

 In Fig. \ref{suppl:fig:suppl_mass_time} we plot the  average flow rate $\dot{Q}/\dot{Q}_{\infty}$ versus obstacle tip height $h$. Here, the flow rate is a measure of the  average mass that  flows through the hopper   during a single discharge event, i.e., 
   $\dot{Q} = \langle m /\tau \rangle $ and $\dot{Q}_{\infty} = \langle m_{\infty} / \tau_{\infty} \rangle $. Figure \ref{suppl_mass_vs_time_obstacle} shows the  variation of the discharged mass $m$  with the   discharge time $\tau$   for repeated experiments under identical experimental conditions, $h=4.4d$ and $D=0.5d$. The  slope of the straight line fitted to this data  is the flow rate $\dot{Q}$.   The normalization factor $\dot{Q}_{\infty}$ represents the average flow rate from the silo without obstacle.  For small values of the obstacle height the steric hindrance provided by the obstacle promotes jamming, i.e. $\dot{Q}_{h\rightarrow 0} \approx 0$. The discharge rate increases with $h$. In experiments, it saturates to $\dot{Q}_{\infty}$, however, in simulation we find an enhancement in the flow rate in the vicinity of $h_{pk}$.

   The data obtained in the experiments are relatively insensitive to the three different clamping conditions of the rod that we tried.   Figure \ref{suppl:fig:suppl_fig_clamping} depicts the three clamping conditions: (a) suspended, (b) clamped and (c) hinged.  The majority of the measurements in the article are taken with a suspended rod because it allows for self-centering rather than relying on perfect lateral positioning of the rod. In Fig. \ref{suppl:fig:suppl_mass_time}  we compare the discharge mass and the duration for different clamping conditions.

We have shown in the main text the height of the average height of the clog $\langle h_{clog} \rangle$   goes to 0  for $h>h_{pk}$. In  Fig. \ref{suppl:fig:suppl_arch_hist}  we plot the probability distribution of $h_{clog}$ for different values of $h$. For $h \le h_{pk}$, the distribution is bimodal. This distribution peaks at $h_{clog}=0$  for values of $h$ that is greater than $h_{pk}$. When the tip of the obstacle is near the orifice, it can influence the formation of the arches that cause the clogging. However, when $h>> h_{pk}$ the obstacle has little effect on the clogging, and the clog happens mainly at the orifice.

 \begin{figure}
   
    \includegraphics[width=1\textwidth]{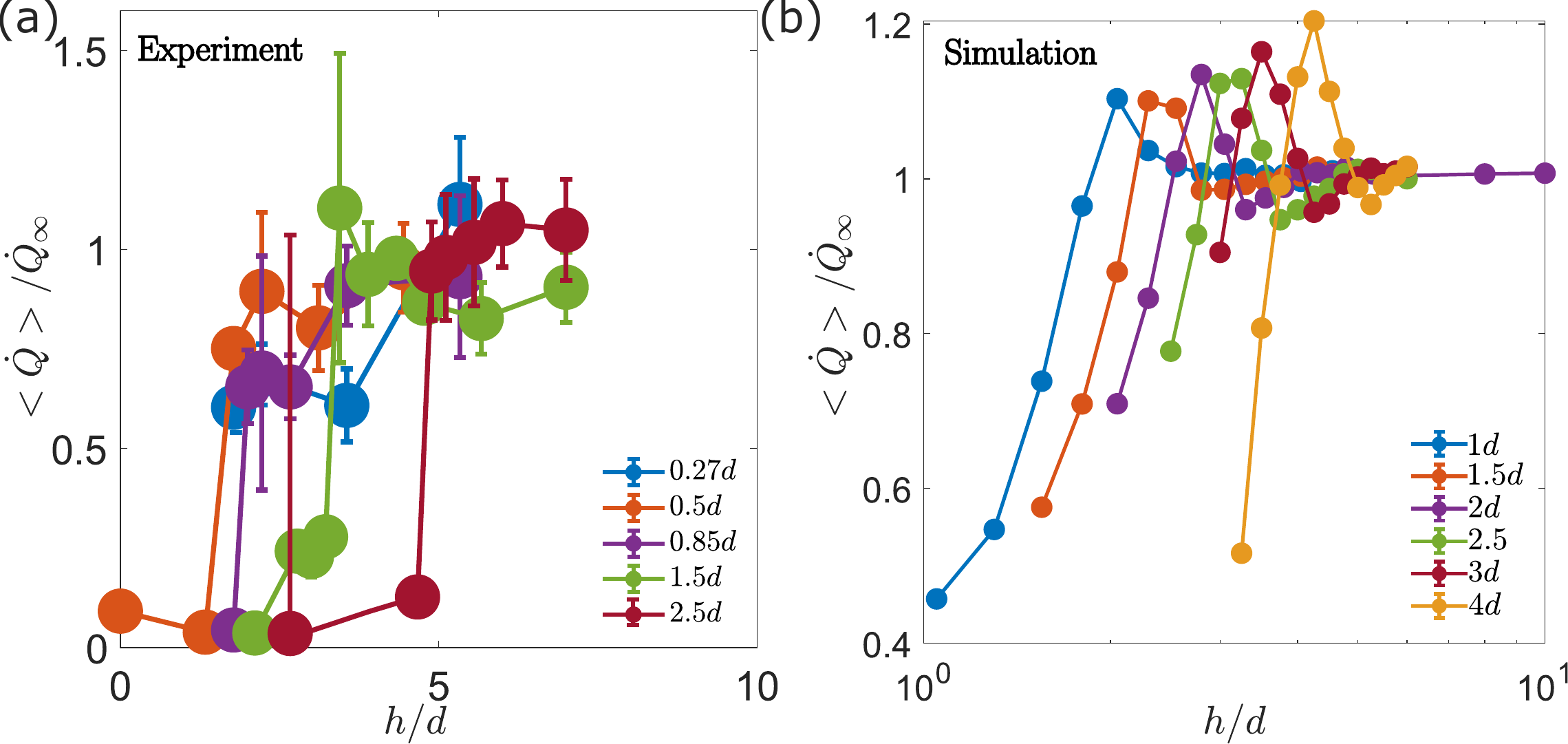}

   \caption{ \textbf{Flow Rate:} Normalized average flow rate $\dot{Q}/\dot{Q}_{\infty}$ versus obstacle tip height $h$. Here, the flow rate is a measure of the  average mass that  flows through the hopper   during a single discharge event, i.e., 
   $\dot{Q} = \langle m /\tau \rangle $ and $\dot{Q}_{\infty} = \langle m_{\infty} /\tau_{\infty} \rangle $.The left panel (a) corresponds to experiments and the right panel (b) to the simulations. The simulations show a small enhancement near the optimal $h$ as has been observed in 2 dimensions.}   
   \label{suppl:fig:suppl_Flow_rate}
 \end{figure}

\begin{figure}
    \centering
    \includegraphics[width=1\linewidth]{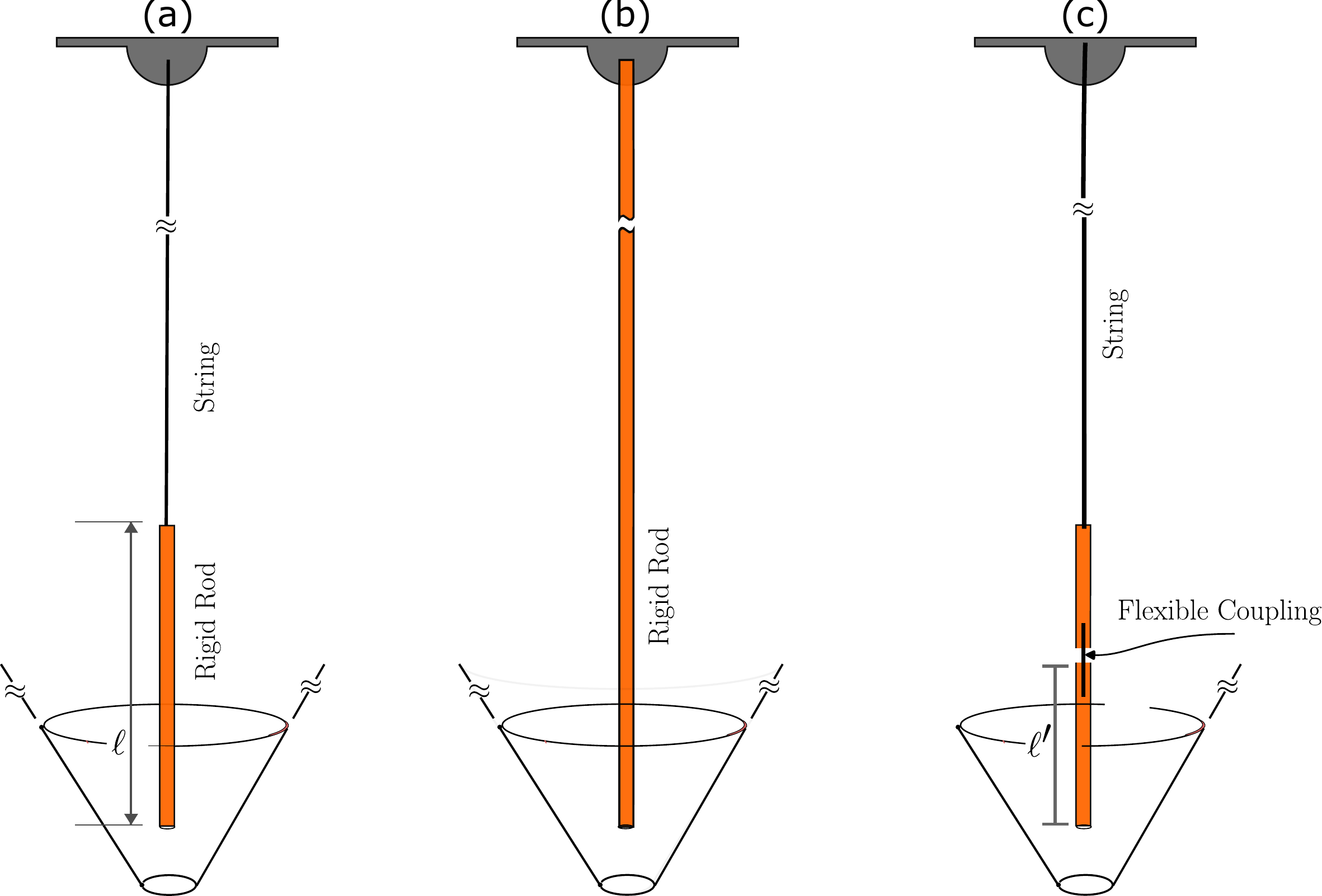}
    \caption{The rod is clamped in three different ways.  (a) Suspended: The upper tip of the rod is suspended from an inextensible string. The string is attached to a support above. During the flow, the rod remains centered and barely moves.   (b) Clamped: The rod is rigidly clamped at the top. (c) Hinged: To allow bending near the outlet of the hopper, we connected two rod segments by a flexible joint and suspended them from a string as in condition (a).} %
    \label{suppl:fig:suppl_fig_clamping}
\end{figure}

  \begin{figure}
  \includegraphics[width=1\textwidth]{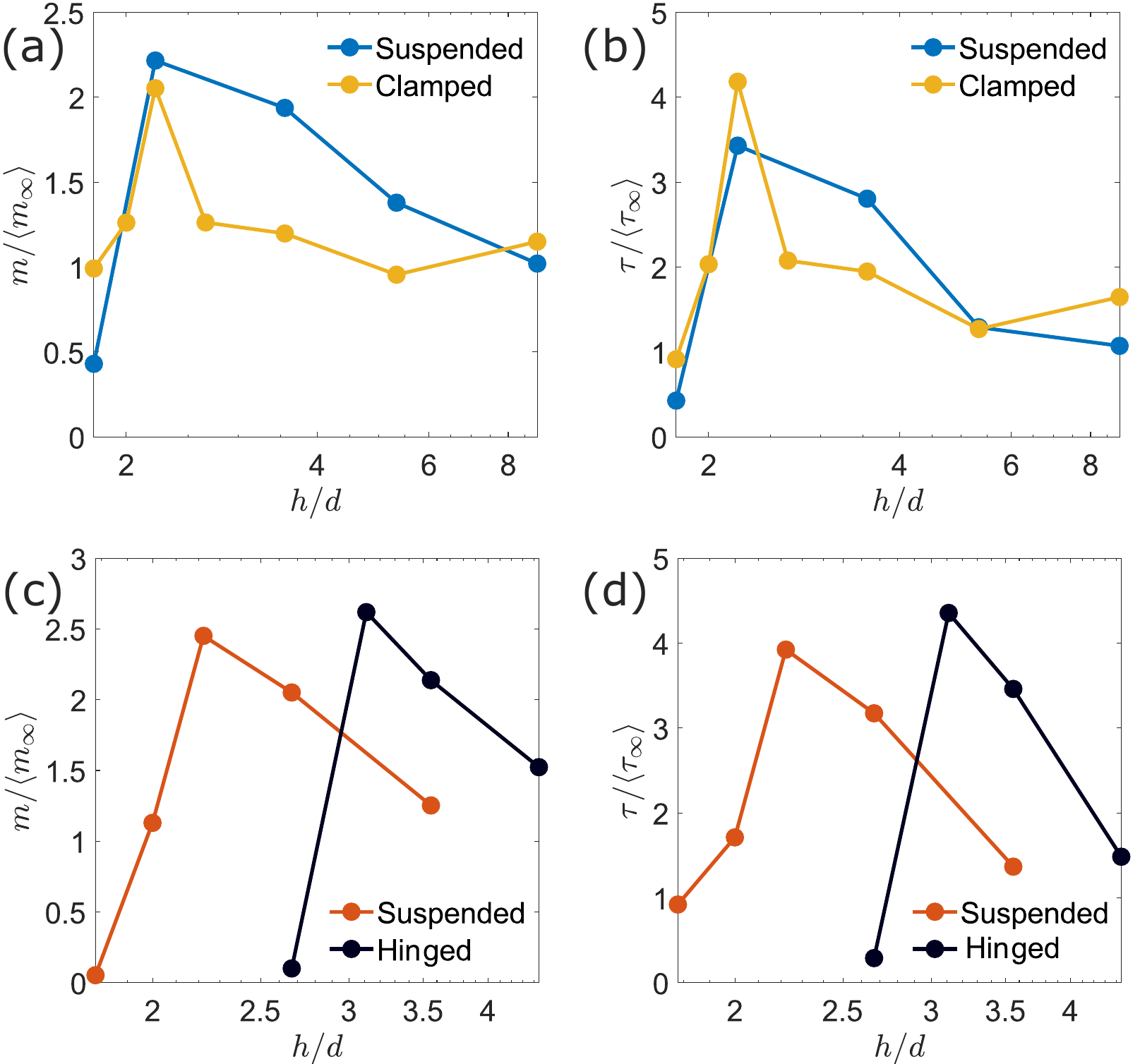}
   \caption{\textbf{Effect of clamping condition} Normalized (a) discharge and (b) duration for rod diameter d$=0.6$mm for different boundary conditions of the rod. (c) discharge and (d) duration for rod diameter d$=1.8$mm for different boundary conditions of the rod. If the rod is carefully centered, clamping conditions do not appear to have a significant effect.}
    \label{suppl:fig:boundary_conditions}
  \end{figure}

 \begin{figure}[t]
    \centering
    \includegraphics[width=1\linewidth]{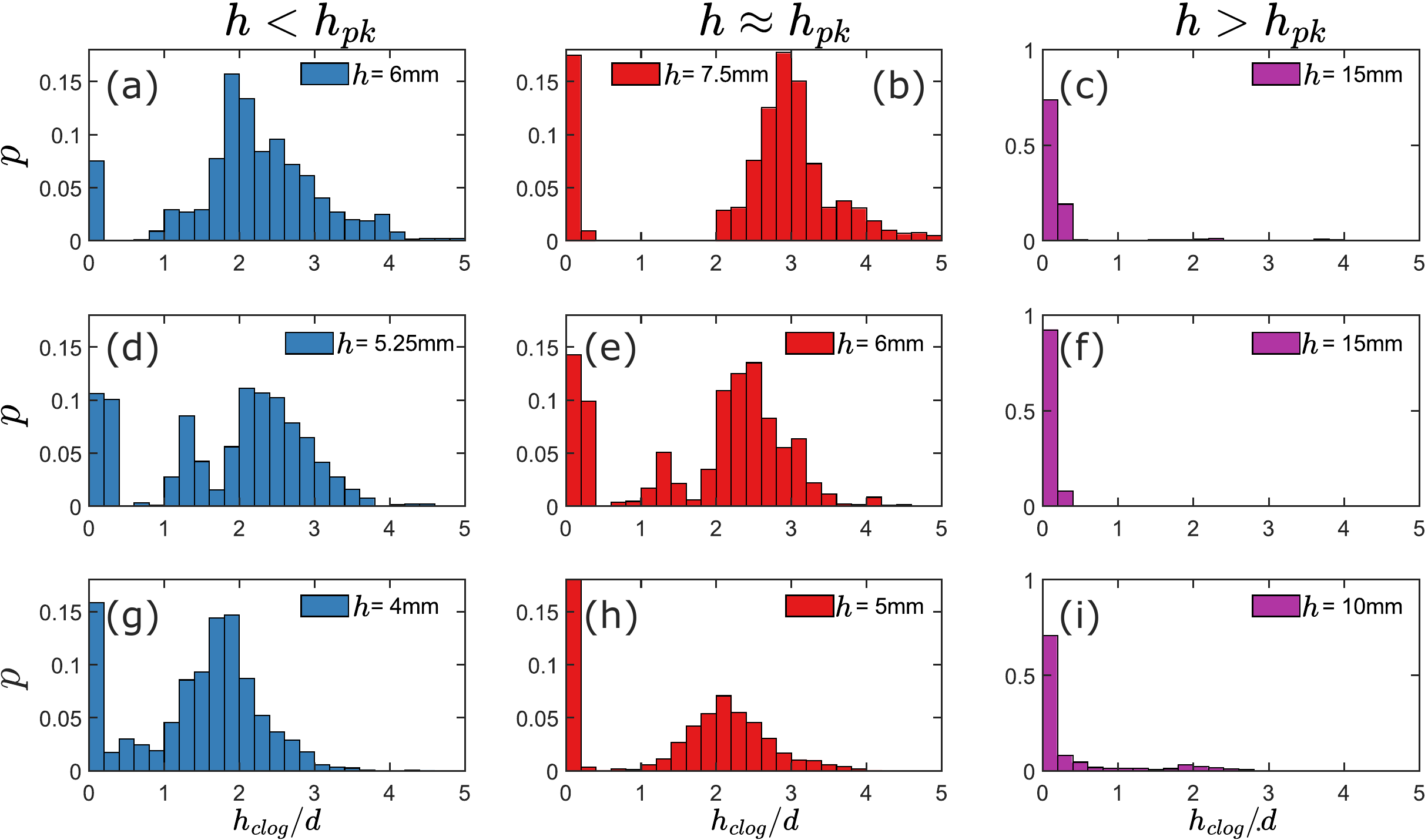}
    \caption{\textbf{Arch height distribution:}
    Probability distributions of arch height $h_{clog}$. 
The rows show the distribution for different obstacle diameters D=0.6 mm, D=1.5 mm, and D=3 mm respectively. The columns show data for $h$ below, at, and above $h_{pk}$. For all rod diameters, the distribution is bimodal when $h<h_{pk}$, but the arch position moves sharply to the outlet $h_{clog}=0$ for $h>h_{pk}$.   }
    \label{suppl:fig:suppl_arch_hist}
     \end{figure}

\clearpage

\newpage

\section{Layering at the walls: from the simulations}
\begin{figure} [htb]
    \centering
    \includegraphics[width=.75\linewidth]{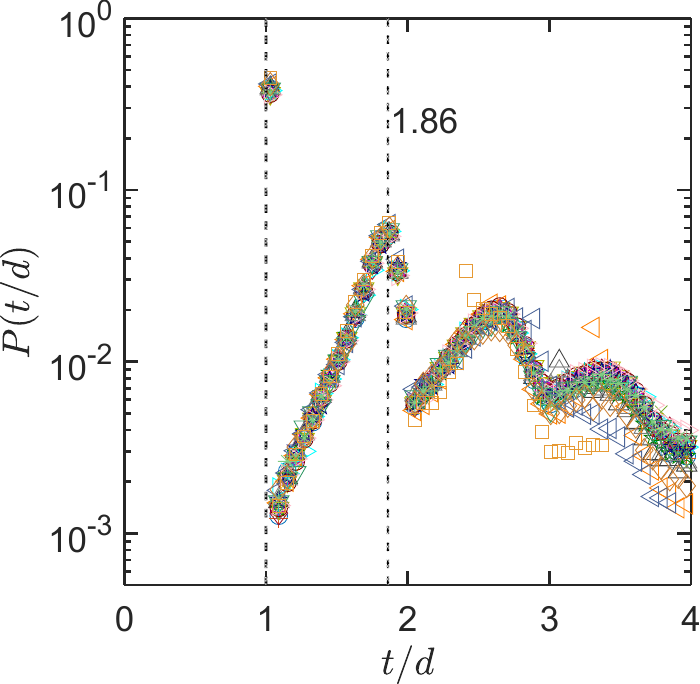}
    \caption{Probability distribution $P(t/d)$ from simulations of the distance $t$ of the particles'  furthermost point  from the hopper's walls. Only the particles in the conical part of the hopper are analyzed. Data for all opening sizes and obstacle heights are included. The first peak in $P(t/d)$ is at $t/d=1$ and the second peak is at $t/d=1.86$.}
    \label{suppl:fig:suppl_t_dist}
\end{figure}

 The simulations show that particles are arranged in layers along the hopper walls, motivating a geometric understanding of clogging.  The layers are best captured in the probability distribution of the  distance, $t = d/2 + r_{wall}$, of the furthermost edge  of the particles from the inner surface of the hopper. Here,  where $r_{wall}$ is the distance of the center of the  particle from the inner surface of the hopper . This distribution is shown in Fig. \ref{suppl:fig:suppl_t_dist}. The peaks at $t= d$ and $t= 1.86 d$  correspond to the first and second layer of particles. This observation motivates a 2D geometric model described in the next section.

\newpage
 \section{Geometric model and justification}
\begin{figure}[!ht]
\includegraphics[width=0.7\textwidth]{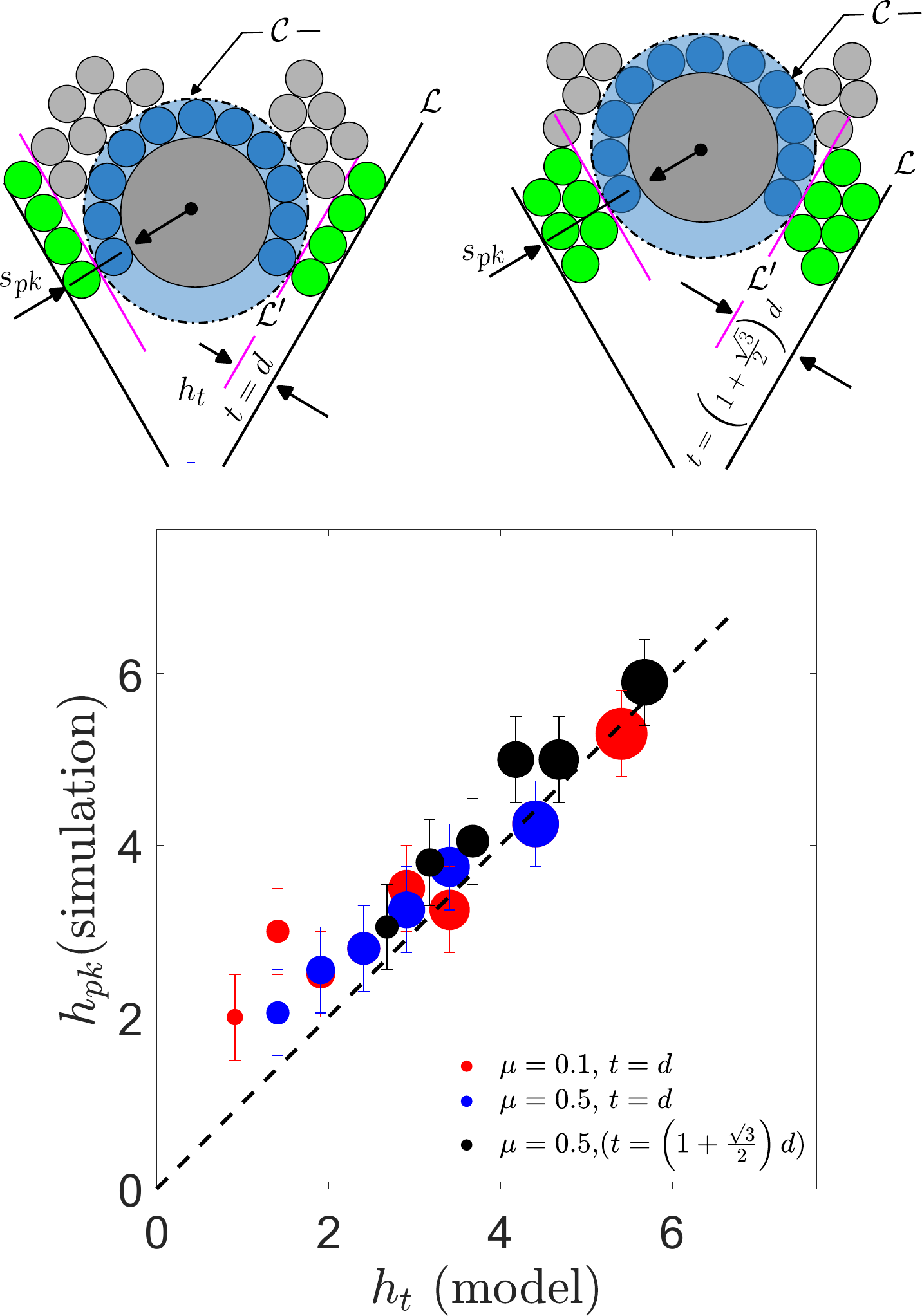}
    \caption{\textbf{Model: } The figure shows the variation of the $h_{pk}$  as obtained from the simulations concerning the corresponding  value of 
    $h_{t}$ obtained from Eqn. \ref{Eqn:model}. The dotted line is the Eqn. $h_{pk}=h_{t}$. The symbol size scales with obstacle diameter $D$.}
    \label{suppl:fig:suppl_model}
    \end{figure}

In Fig. \ref{suppl:fig:suppl_model} we show a schematic 2D version of the hopper, with discs for the obstacle and particles. The layer of particles in contact with the obstacle forms part an arc of a circle $\mathcal{C}$ (diameter $D+d$). At $h = h_{pk}$, the arch is tangent to the outer envelope $\mathcal{L}':y=\sqrt{3}x - (-2d+ \sqrt{3}a/2)$ of the layer of particles in contact with the wall. This amounts to finding the value of $h$ where $\mathcal{L}: y=\sqrt{3}(x-a/2)$, where $\mathcal{L'} \parallel \mathcal{L}$.
From geometry, we obtain :
\begin{equation}
h_{model} = \left( \frac{D}{2} + 1 \right) \sec(\theta) - \left( \frac{a}{2} \right) \cot(\theta) - t \cos(\theta) \text{cosec}^2(\theta).
\label{Eqn:model}    
\end{equation}
Here $t$ is the  distance between the layers $\mathcal{L}$ and $\mathcal{L}'$.  
The observed $ h_{pk} \approx  h_{t}$, computed from this model  \ref{suppl:fig:suppl_model}.

\paragraph{\textbf{Simulation:}}
The shortest distance $ s_{s}$ between the surface of a sphere with radius $ d/2 $, centered at $ (0,h) $, and the wall $ \mathcal{L} $ of the hopper is given by:
\[
s_{s}(h) = \frac{\lvert2h + \sqrt{3}a - 2d\rvert}{4}
\]
The difference between short distance for a given height is given by \[s_{c}-s_{s}=(2-\sqrt{3})d/4\]

\paragraph{\textbf{Experiment:}}
Modeling the experiment is more complicated, as the particles are not monodisperse, and layering will not be as distinct. However, for completeness, the shortest distance $s_{c}$  of the point $(d/2,h)$ on a cylinder  and the  wall $\mathcal{L}$ of the  hopper is given by:
\[
s_{c}=\frac{\lvert2h-\sqrt{3}(d-a)\rvert}{4}
\]

\clearpage

\bibliographystyle{apsrev4-1}
\bibliography{hopper}